\documentclass[amsmath,amssymb,pra,final,showpacs,twocolumn]{revtex4-2}
 
\usepackage{bm,hyphenat,xspace}
\usepackage{graphicx,epsfig}

\newcommand {\mct}{\mathcal{T}}
\newcommand {\mcu}{\mathcal{U}}

\newcommand{\He}{${}^4\mathrm{He}$ }

\begin{document}

\title {Second excited state of  ${}^4\mathrm{He}$ tetramer}
  
\author{A.~Deltuva} 
\affiliation
{Institute of Theoretical Physics and Astronomy, 
Vilnius University, Saul\.etekio al. 3, LT-10257 Vilnius, Lithuania
}

\received{October 13, 2025} 

\begin{abstract}
  The four-boson universality  suggests the existence of the second excited tetramer state in a
   system of cold  ${}^4\mathrm{He}$ atoms. It is not bound but could be seen as a resonance in the
   atom-trimer scattering. This process is rigorously calculated  using the  momentum-space  transition operator
   framework with two realistic interatomic  potentials. The $S$-wave phase shift and cross section show a
   resonant behavior  below the excited trimer threshold, but there are sizable nonresonant contributions
   from $P$ and $D$ waves as well. The position and width of the resonant state is determined, for the latter
   significant finite-range effects are found.
\end{abstract}

 \maketitle

\section{Introduction \label{sec:intro}}

The few-body systems of atomic  \He atoms are among the best known examples for the natural realization  of
the Efimov physics \cite{efimov:plb,kunitski:15a}.
The two-atom scattering length $a$ is large compared to the range of the interatomic van der Waals interaction 
$r_{\rm vdW}$
and the effective range, supporting only one very  weakly  bound two-body state in the $S$-wave, the dimer
with the binding energy $B_2 \sim  \hbar^2/m_a a^2 \sim 1$ mK, $m_a$ being the mass of the atom.
Consequently \cite{braaten:rev,naidon:rev,kievsky:21arnps},
several weakly bound states should exist also in the systems
consisting of three and four \He atoms. Indeed, a number of calculations using realistic interaction models
\cite{aziz,przybytek:10a} predicted two states of the trimer
\cite{blume:00a,barletta:01a,lazauskas:he,suno:08a,kolganova:09a,roudnev:11a,das:11a,hiyama:12a,deltuva:15g},
a deep one with the binding energy $B_3 \sim 100 \, B_2$,  and a shallow one close to the atom-dimer threshold,
$B_2 < B_3^* < 2\, B_2$.
Similarly, two bound states were predicted also in the four-atom system
 \cite{blume:00a,lazauskas:he,das:11a,hiyama:12a,deltuva:22a},
one deep and one shallow, close to the atom-trimer threshold. In the latter case there are significant
differences in predicted binding energies, with only three most advanced calculations 
\cite{lazauskas:he,hiyama:12a,deltuva:22a} being in a reasonable agreement. 
The above-mentioned tetramer states are associated with the trimer ground state, their binding energies
roughly being $B_4 \sim 4.5 \, B_3$ and $B_4^* \sim 1.01 \, B_3$, respectively.

\begin{figure}[!]
\begin{center}
\includegraphics[scale=0.4]{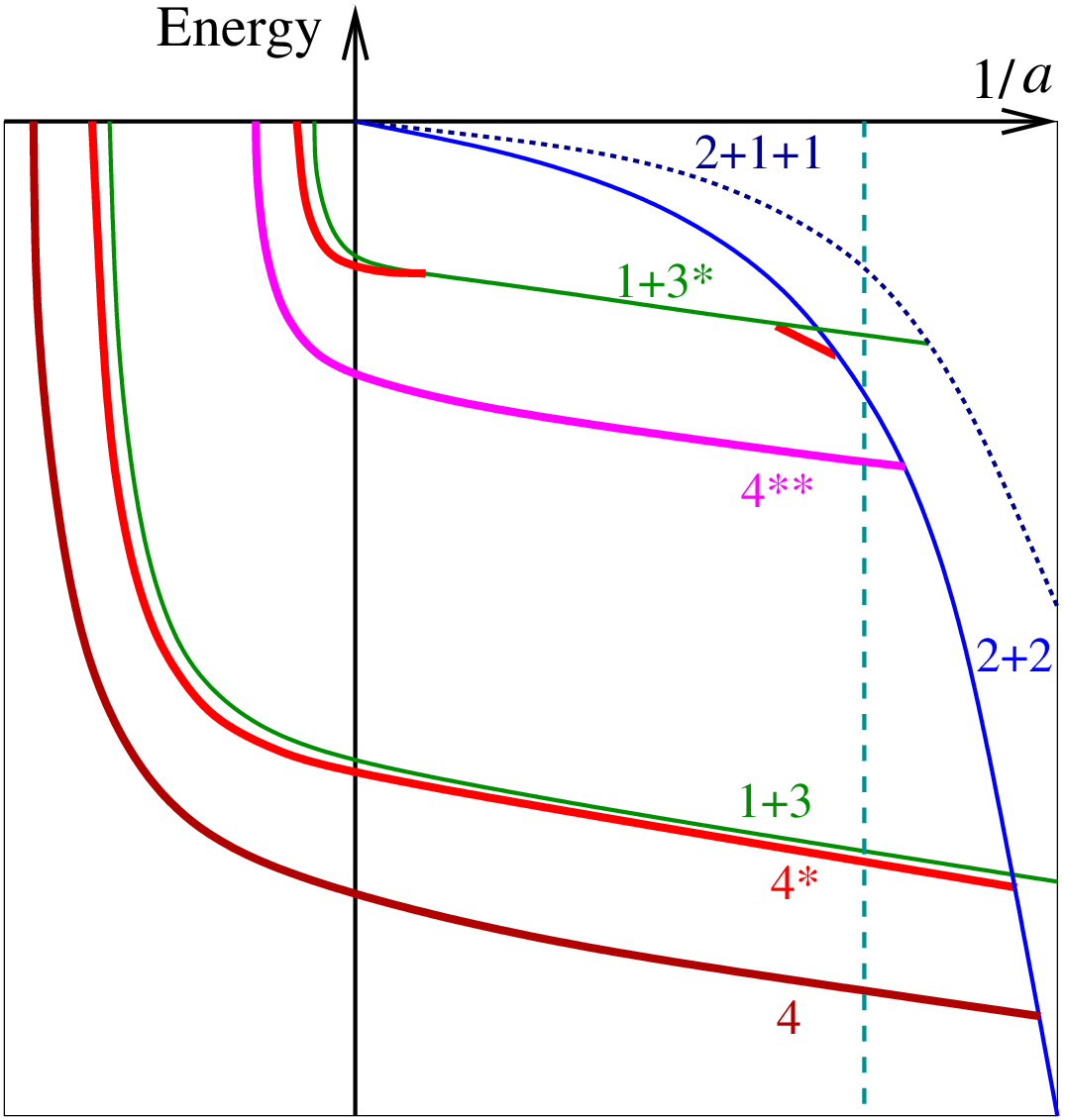}
\end{center}
\caption{\label{fig:spectrum} (Color online)
Schematic Efimov plot for the system of four \He atoms, i.e., the energy levels and thresholds 
as functions of the inverse two-atom scattering length. Only two families of states are shown.
The energies of tetramers are displayed by thick curves;
those labeled 4, 4* and 4** correspond to the ground, first  and second excited states, respectively.
The thin curves correspond to few-cluster thresholds.
The vertical dashed line labels the physical \He point.
}
\end{figure}

In the ideal Efimov scenario there should be two tetramer states associated with each trimer state
\cite{hammer:07a,stecher:09a,deltuva:10c}.
The tetramers associated with the excited states of the trimer are not truly bound states,
but resonant-like unstable bound states (UBS) residing in the continuum \cite{res_cpl,deltuva:11a}.
Their properties have been calculated with high accuracy using simple separable potentials
\cite{deltuva:10c,deltuva:11a,deltuva:12d}. The results of Ref.~\cite{deltuva:12d} suggest that for 
$B_3^*/B_2$ values corresponding to a realistic \He atomic system there is only one associated physically
observable tetramer, the deeper one. When increasing $1/a$ from the unitary limit toward the
physical point the shallow tetramer turns into a virtual state through the  dimer-dimer threshold and moves
far away from the physical region,  therefore becoming unobservable.
The schematic representation of energy levels and thresholds in the \He four-atom system
is displayed in Fig. \ref{fig:spectrum} 

For a realistic system of \He atoms the second excited tetramer state has never been calculated. 
There are two main difficulties related:  (i) the state is not a bound state
but the atom-trimer scattering state located high
above the two-cluster threshold, while the absolute majority of \He four-body calculations are for bound states
\cite{blume:00a,das:11a,hiyama:12a} or in the vicinity of the atom-trimer threshold
\cite{lazauskas:he}; (ii)  the 
realistic interaction between two \He atoms has weakly attractive van der Waals tail 
and very strong  short-distance repulsion
\cite{aziz,przybytek:10a,PhysRevLett.119.123401}, that make the numerical solution highly challenging
and very sensitive to fine details,  especially for spatially extended  states.
These difficulties have been overcome recently using 
Alt, Grassberger, and Sandhas (AGS) equations \cite{grassberger:67}
for the four-particle transition operators together with the ``softening and extrapolation'' method 
\cite{deltuva:22a}, where the strength of the short-range repulsion was gradually reduced,
enabling accurate solutions of dynamical equations,
with subsequent extrapolation of the results back to the original potential.
Solving four-body equations in the momentum space, Ref.~\cite{deltuva:22a}
obtained not only tetramer binding energies but also atom-trimer and dimer-dimer scattering lengths.

The aim of the present work is to apply  the momentum-space method 
from Ref.~\cite{deltuva:22a} to the atom-trimer scattering calculation near the excited trimer threshold
and extract the properties of the resonant state, the second excited state of the tetramer.
Further conclusions will be drawn regarding its observability, model dependence, and the importance of the
finite-range effects.

 Section II shortly recalls the desctiption of the four-body scattering using transition operators
 together with the essential aspects of calculations. Section III presents results for the
 atom-trimer  scattering in the vicinity of the resonant state.
Discussion and conclusions are collected in Sec. VI.

\section{Atom-trimer scattering equations \label{sec:eq}}

The  AGS equations \cite{grassberger:67} can be considered an integral version of the 
Faddeev-Yakubovsky equations \cite{yakubovsky:67}. While the latter are formulated for 
wave-function components, the AGS equations 
\begin{subequations} \label{eq:U}
\begin{align}  
\mcu_{11}  = {}&  P_{34} (G_0  t  G_0)^{-1}  
+ P_{34}  U_1 G_0  t G_0  \mcu_{11}   + U_2 G_0  t G_0  \mcu_{21} , 
\label{eq:U11} \\  
\mcu_{21}  = {}&  (1 + P_{34}) (G_0  t  G_0)^{-1}  
+ (1 + P_{34}) U_1 G_0  t  G_0  \mcu_{11} , \label{eq:U21} 
\end{align}
\end{subequations}
relate the four-particle transition operators $\mcu_{\beta\alpha}$ to the two-particle transition operator
\begin{equation} \label{eq:t}
t= v + v G_0 t
\end{equation}
and  3+1 and 2+2 subsystem transition operators
\begin{equation} \label{eq:U3}
U_{\alpha} =  P_\alpha G_0^{-1} + P_\alpha  t G_0  U_{\alpha}.
\end{equation}
The above equations assume the bosonic symmetry, $v$ is the two-particle potential,
\begin{equation} \label{eq:g0}
G_0 = (E+i0-H_0)^{-1}
\end{equation}
 is the free four-body resolvent at the available system energy $E$, and
$H_0$ is the kinetic energy operator.
The Greek subscripts $\alpha = 1$ (2) label the 3+1 (2+2) clustering,
$P_1 =  P_{12}\, P_{23} + P_{13}\, P_{23}$ and $P_2 =  P_{13}\, P_{24} $,
where $P_{ab}$ is the permutation operator of  particles $a$ and $b$.

The AGS equations (\ref{eq:U}) are solved in the momentum-space partial-wave representation
 $|k_x k_y k_z [(l_x l_y)j l_z] {JM} \rangle_\alpha$,  becoming an integral equation system
 with three continuous variables  $k_x, k_y, k_z$ that are the  magnitudes of  Jacobi momenta \cite{deltuva:07a}. 
After their discretization  a large system of linear algebraic equations is obtained.
The   orbital angular momenta $l_x, l_y, l_z$ are associated with the Jacobi momenta $k_x, k_y, k_z$.
They  are coupled to the total angular momentum ${J}$ with the projection ${M}$,
whereas $j$ is an intermediate angular momentum.
More details on the solution methods can be found in  Refs.~\cite{deltuva:22a,deltuva:07a}.

Partial-wave amplitudes for the atom-trimer scattering  are calculated as on-shell
matrix elements
\begin{equation} \label{eq:ampl}
\mct_J(E) = 3 \langle \phi p {J} | \mcu_{11}| \phi p  {J}  \rangle
\end{equation}
of the transition operator $\mcu_{11}$ between
the channel states $| \phi p {J} \rangle$ \cite{deltuva:07a}.
These states are direct products of the trimer ground state
Faddeev amplitude $\phi$ and a free wave for the atom-trimer motion with
total angular momentum $J$ and 
relative two-cluster momentum $p$, related to the total energy $E = -B_3 + p^2/2\mu$,
$\mu=3m_a/4$ being the reduced mass. The factor 3 in Eq.~(\ref{eq:ampl}) stems from the symmetrization.

 This scattering amplitude is related to the single-channel 
$S$-matrix and scattering phase shift
 $\delta_{J}$ in a standard way, i.e., 
\begin{equation} \label{eq:S}
 \mathcal{S}_{J}(E) =  e^{2i\delta_{J}(E)} = 
1 - 2i \pi \mu p \mct_{J}(E).
\end{equation}
Note that parity of the channel states is given by $\Pi = (-1)^J$ since the trimer has total spin zero.

\section{Results  \label{sec:res}}

\begin{table}[!]
  \caption{Calculated dimer and trimer binding energies in units of mK.
    The numerical error is at most 1 in the last significant digit.
    The last line contains LM2M2 results obtained with $l_x, l_y \le 8$.
}
\label{tab:b}
\centering
\begin{ruledtabular}
\begin{tabular}{{l}*{3}{r}}
  & $B_2$  & $B_3^*$ & $B_3$  \\ \hline
  PCJS  & 1.6125 & 2.646 & 131.58 \\
  LM2M2 & 1.3094 & 2.277 &  126.30 \\
  LM2M2(8) & 1.3094 & 2.278 & 126.50 \\
\end{tabular}
\end{ruledtabular}
\end{table}

The results are obtained using two realistic interatomic \He potentials.
The older  parametrization  LM2M2 by Aziz and Slaman \cite{aziz} is widely used for benchmarks,
while that by Przybytek et al. (PCJS)
\cite{PhysRevLett.119.123401} is among the most recent and sophisticated ones.
They differ by about 20\% in $B_2$ and $B_3^*$ predictions, thereby allowing to explore the
model dependence. As all realistic \He potentials, they have  weakly attractive van der Waals tail 
and  strong repulsion at short distances; this difficulty is reliably handled using the
``softening and extrapolation'' method from  Ref.~\cite{deltuva:22a}.
The calculations include partial waves  with $l_x, l_y, l_z \le 6$, where
the trimer excited (ground) state binding energies are converged  better than within 0.1\% (0.2\%).
The $\hbar^2/m_a=12.11928$ K\AA$^2$ value recommended in Ref.~\cite{roudnev:11a} is adopted.
Table \ref{tab:b} collects the values for dimer and trimer binding energies.

\begin{figure}[!]
\begin{center}
\includegraphics[scale=0.64]{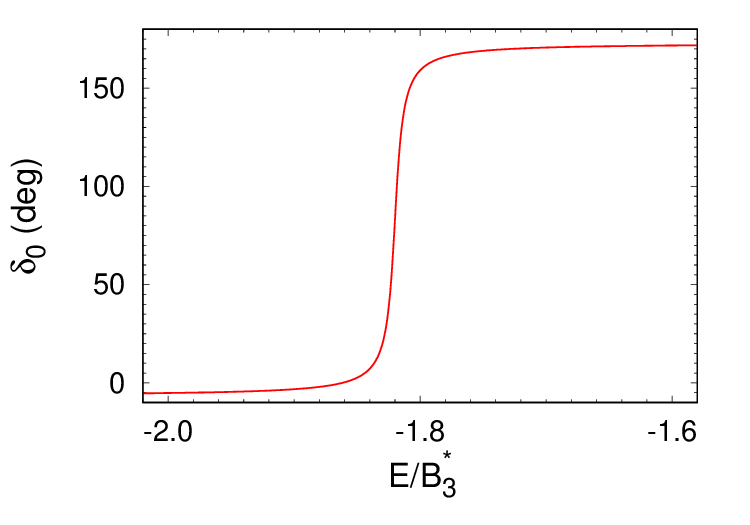}
\end{center}
\caption{\label{fig:phase} (Color online)
Energy dependence of the
  $J=0$ phase shift for the LM2M2 interaction model in the vicinity of the
  second excited tetramer state.
}
\end{figure}

Solving the atom-trimer scattering equations (\ref{eq:U}) and calculating the phase shift
(\ref{eq:S}) reveals a very clear resonant behavior in the $J=0$ state around the energy
$E \approx -1.8\,B_3^*$. It turns out that in this region the phase shift can be very well
parametrized as a sum of slowly-varying background and resonant contributions \cite{taylor:book},
i.e.,
\begin{equation} \label{eq:ph}
  \delta_{J}(E) = \varphi_0 + \varphi_1\, \frac{E}{B_3^*} +
  \arcsin{\left[ \frac{\Gamma/2}{\sqrt{(E+B_4^{**})^2 + (\Gamma/2)^2}}\right] }.
\end{equation}
Here $B_4^{**}$ parametrizes the position of the resonance below the threshold of four free particles,
while $\Gamma$ is its width, i.e., in the complex energy plane the pole of the four-particle
$S$-matrix and transition operators is located at $-B_4^{**} - i\Gamma/2$.
The phase shift for the LM2M2 potential is shown in Fig.~\ref{fig:phase},
where the values of parameters in Eq.~(\ref{eq:ph}) are
$\varphi_0 = -7.8(1)$ deg, $\varphi_1 = -0.51(3)$ deg,
$B_4^{**}/B_3^* = 1.82(2)$, and $\Gamma/B_3^* = 0.010(1)$. 
Since the second excited tetramer state is associated with the first excited trimer, 
$B_3^*$ is chosen as the most natural energy scale in Eq.~(\ref{eq:ph}). Nevertheless, one could use
$B_3$ as well taking the ratio from Table \ref{tab:b}.

\begin{figure}[!]
\begin{center}
\includegraphics[scale=0.64]{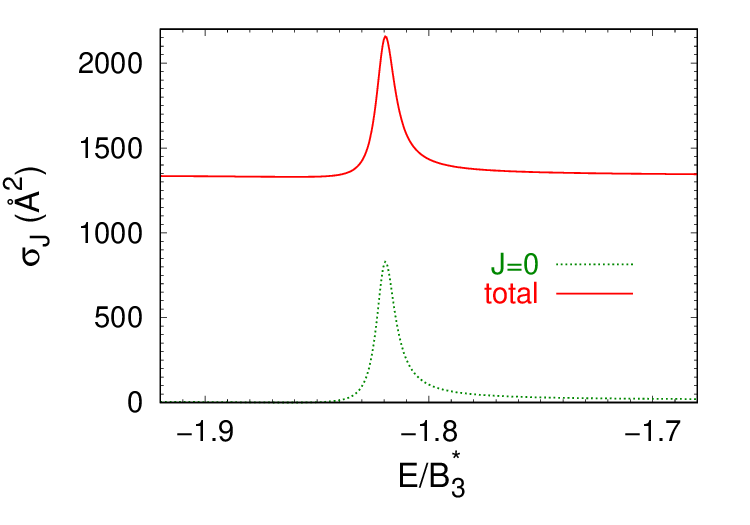}
\end{center}
\caption{\label{fig:cs} (Color online)
The total cross section and its  $J=0$ contribution 
as functions of energy   calculated using the
  LM2M2 interaction model in the vicinity of the second excited tetramer state.
}
\end{figure}

The contribution of the $J=0$ state to the total cross section
\begin{equation} \label{eq:cs}
  \sigma_{J}(E) = 4\pi (2J+1)\, \frac{\sin{\delta_J(E)}}{p^2}
\end{equation}
is shown in Fig.~\ref{fig:cs}, exhibiting a sharp resonant peak, where the cross section rises by a factor
of 100 as compared to the background.

However, one has to keep in mind that the considered regime is relatively high above the
atom-trimer threshold and therefore other partial waves with $J>0$ contribute as well.
They do not show any resonant behavior, the corresponding phase shifts and cross sections
do not change significantly in the considered regime. Their values at the $J=0$ resonance energy
are collected in Table \ref{tab:s}. It appears that $J=1$ and $J=2$ waves at this energy
have appreciable values of phase shifts and cross sections, in the $J=1$ case  even exceeding
the resonant contribution. In contrast, in  $J=3$ and higher waves  the phase shifts in
absolute values are below 1 deg, with entirely insignificant contributions to the total
cross section, below 1 $\mathrm{\AA}^2$.

\begin{table}[!h]
  \caption{Phase shifts and contributions to the total cross section at the energy
    corresponding to the $J=0$ cross section peak. Results are obtained using the
    LM2M2 potential.
}
\label{tab:s}
\centering
\begin{ruledtabular}
\begin{tabular}{*{3}{r}}
  $J$ & $\delta_J$ (deg) & $\sigma_J \, (\mathrm{\AA}^2)$ \\ \hline
  0 & 90.0(1) & 830(1) \\
  1 & -41.1(1) & 1076(3) \\
  2 & 14.3(1) & 253(2) \\
\end{tabular}
\end{ruledtabular}
\end{table}

The calculations using the PCJS potential \cite{PhysRevLett.119.123401}
yield very similar results. The extracted 
$B_4^{**}/B_3^*$, and $\Gamma/B_3^*$ values are collected in Table~\ref{tab:b4}.

\begin{table}[!]
  \caption{Position and width of the second excited tetramer state in ratio to the
    excited trimer binding energy. The universal results correspond to the
    $B_3^*/B_2 = 1.74$ ratio of  the LM2M2 potential.
}
\label{tab:b4}
\centering
\begin{ruledtabular}
\begin{tabular}{{l}*{2}{l}}
  &  $B_4^{**}/B_3^*$ & $\Gamma/B_3^*$ \\ \hline
  PCJS  & 
  1.79(2) & 0.009(1) \\
  LM2M2 & 
  1.82(2) & 0.010(1) \\
  Universal & 1.888(1) & 0.0048(1) \\
\end{tabular}
\end{ruledtabular}
\end{table}

\section{Discussion and conclusions \label{sec:sum}}

The universal Efimov scenario predicts the existence of the second excited tetramer state in a
realistic system of cold \He atoms. Being well above the atom plus ground-state trimer threshold,
this state is an unstable bound state in the continuum that can be seen as a resonance in 
atom-trimer collisions. The difficulties in handling the four-particle scattering  with van der Waals
plus strongly repulsive potentials have been overcome by using the  momentum-space  transition operator
framework together with the ``softening and extrapolation'' method \cite{deltuva:22a}.

The resonant behavior is very  clearly pronounced in the $J=0$ state. However, higher total angular
momentum states significantly contribute to the nonresonant cross section, such that the increase of the
total atom-trimer cross section at the peak is about 60\% as shown in Fig.~\ref{fig:cs}. 
This should be observable, although
less dramatic than peaks predicted for the recombination at threshold \cite{stecher:09a,deltuva:12a}, 
dominated by the single $J=0$ state.

Another important question is the size of finite-range effects for realistic interatomic \He potentials,
as compared to the strictly universal (zero-range) Efimov scenario. The latter results can be obtained
from Ref.~\cite{deltuva:12d}, and are listed in Table~\ref{tab:b4}.
 The universal system with the same binding energy ratio
 as the LM2M2 potential would have the second excited tetramer state
characterized by $B_4^{**}/B_3^* = 1.888(1)$ and $\Gamma/B_3^* = 0.0048(1)$.
Comparison with direct  LM2M2 results 
reveals quite sizable finite-range effects, especially for the resonance width that 
is twice as large as the universal prediction. 
The model dependence, estimated as the difference between LM2M2 and  PCJS potential results,
 is considerably less significant. However,  it becomes
 more visible for the resonance position in
 absolute values, i.e.,  $B_4^{**} = 4.14(4)$ mK  for LM2M2
 and $ 4.74(5)$ mK for PCJS, while the width is almost the same,
 $\Gamma = 0.023(2)$ mK  and $ 0.024(2)$ mK, respectively.

Thus, although partially masked by higher waves, the second excited \He tetramer state should be observable
in  atom-trimer collisions as  quite sharp resonance that is broadened due to finite range effects.

\vspace{1mm}



\end{document}